# Phase dependence of the Thermal Memory Effect in Polycrystalline Ribbon and Bulk $Ni_{55}Fe_{19}Ga_{26}$ Heusler Alloys

A. Vidal-Crespo[1], A.F. Manchón-Gordón[2,*], J.M. Martín-Olalla[1], F.J. Romero[1], J.J. Ipus[1], M.C. Gallardo[1], J.S. Blázquez[1,*], C.F. Conde[1]

[1]Dpto. Física de la Materia Condensada, ICMSE-CSIC, Universidad de Sevilla, P.O. Box 1065, 41080 Sevilla, Spain
[2]Instituto de Ciencia de Materiales de Sevilla, ICMSE CSIC-Universidad de Sevilla, C. Américo Vespucio 49, Sevilla, 41092, Spain

[*]*The corresponding authors e-mails:* alejandro.manchon@icmse.csic.es, jsebas@us.es

**ABSTRACT:** The thermal memory effect, TME, has been studied in $Ni_{55}Fe_{19}Ga_{26}$ shape memory alloys, fabricated as ribbons via melt-spinning and as pellets via arc-melting, to evaluate its dependence on the martensitic structure and the macrostructure of the samples. When the reverse martensitic transformation is interrupted, a kinetic delay in the subsequent complete transformation is only evident in the ribbon samples, where the 14M modulated structure is the dominant phase. In contrast, degradation of the modulated structure or the presence of the $\gamma$ phase significantly reduces the observed TME. In such cases, the magnitude of the TME approaches the detection limits of commercial calorimeters, and only high-resolution calorimeter at very low heating rate (40 mK h$^{-1}$) can show the effect. Following the kinetic arrest and subsequent cooling, the reverse martensitic transformation was completed at several heating rates to confirm the athermal nature of the phenomenon.

**KEYWORDS:** Martensitic transformation, Thermal memory effect, Shape memory alloys, Ni-Fe-Ga Heusler alloys, ultraslow calorimetry

# 1. INTRODUCTION

The properties of shape memory alloys, SMAs, have attracted considerable interest making them promising for innovative engineering and mechanical applications, primarily due to their unique shape memory effect, SME, and superelasticity, SE [1]. These distinct characteristics arise from the martensitic transformation, MT, that these materials exhibit: a solid-state, first-order phase transition. MT involves the coordinated movement of a considerable number of atoms, achieving a growth velocity similar to sound waves [2]. On cooling, a high-temperature austenite phase with high symmetry transforms into a low-temperature martensite phase with reduced symmetry. The heating transformation is thus called reverse MT. In general, the low-temperature martensitic phase exhibits a multidomain microstructure, resulting from the distortion of the crystal lattice. This structural distortion accounts for many of the distinctive characteristics of these transformations.

In addition to showcasing SME or SE, SMAs have demonstrated their capacity to recall not only specific shapes but also the temperatures at which the reverse MT was intentionally halted [3, 4]. Unlike the well-understood SME, the called thermal memory effect, TME, is less clear, and different scenarios have been proposed encompassing the redistribution of accumulated stress [5, 6] or the impact of geometrical constrictions [7]. In this phenomenon, when an intentional arrest occurred at a stop temperature $T_{stop}$ between austenite start, $A_s$, and austenite finish temperatures, $A_f$, i.e. $T_{stop} \in (A_s, A_f)$, the following complete reverse MT shows a kinetic pause close to $T_{stop}$. It has been previously labeled as thermal arrest memory effect [8], or as a reversible step-wise transformation from martensite to austenite [9].

In recent years, various theoretical approaches have been proposed to explain this phenomenon, with particular emphasis on the relationship between TME and factors such as dislocations [10], interface interactions [11], elastic deformation of the martensitic plates [3, 8], and the continuous distribution of stored elastic energy, which relaxes during partial heating cycles [12]. However, there is no consensus regarding the origin of these effects, and their physical interpretation remains controversial. This uncertainty arises primarily because local stress evolution cannot be easily measured, making it difficult to assess its impact on transformation behavior. This is a significant technological challenge, especially since most real SMA actuators typically experience only partial transformation cycles under working conditions [13]. More recently, *in-situ* neutron diffraction experiments have provided a comprehensive description of evolution of thermoelastic strains occurring in SMAs during partial thermal transformation. As a result, TME is expected in multi-domain crystals undergoing structural first-order phase transitions, where accommodation processes lead to strain distribution among domains [14].

While the TME was initially observed in Ti-Ni alloys [8], it has also been detected in Ni-Mn-Ga [15, 16] and Ni-Fe-Ga [17, 18] Heusler alloys. In the latter case, the MT occurs between an austenite phase with a B2 or ordered $L2_1$ structure to either a modulated (seven-layer, 14M, or five-layer, 10M) or a non-modulated ($L1_0$ tetragonal) martensitic structure. The specific structure depends on factors such as the composition and the thermal and mechanical history [19, 20]. The $\gamma$-phase precipitates are usually formed in conventionally produced samples, resulting in enhanced mechanical properties. These precipitates can be also easily induced by thermal treatments in monophasic melt-spun ribbons [21]. This ductile phase does not participate in the martensitic phase transition, and its presence weakens the SME. As the amount of the ductile phase increases, it hinders the movement of martensite variants, reducing the material's ability to recover strain. In this sense, porosity can affect the transformation favoring the movement of the twin boundaries due to a higher free surface as reported for Ni-Mn-Ga [22]. This phenomenon is enhanced in foams [23]. Therefore, the diverse array of microstructures in which MT can be observed calls for an examination of their influence on the TME.

This study systematically explores the TME arising from interrupting reverse MT in different samples of $Ni_{55}Fe_{16}Ga_{26}$ Heusler alloys fabricated as ribbon-shaped specimens, using melt-spinning, and as pellet-shaped samples, using arc-melting. The study encompassed the execution of either a single interruption of the reverse MT or a series of interrupted processes with several stop temperatures. The main goal was to compare how the martensitic structure and the macrostructure of the sample influences the TME. The findings presented in this study complete previous works by some of the authors on Ni-Fe-Ga SMAs, exploring various aspects: i) the influence of pressure on the phase stability and magnetostructural transition of ribbons [20], ii) distinctions in structural and magnetic properties between ribbon and bulk samples [21], iii) the kinetics [24] and isothermal/athermal nature [18] of the MT in ribbons, and iv) the identification of avalanches during the MT in bulk samples using ultraslow calorimetry [25].

## 2. EXPERIMENTAL

The material analyzed in this study is an alloy with a nominal composition of $Ni_{55}Fe_{19}Ga_{26}$ (at. %) fabricated using both melt-spinning and arc-melting techniques. Comprehensive details regarding the preparation procedures and a thorough microstructural and magnetic characterization of both samples are available in reference [21]. Moreover, detailed insights into the martensitic structure at room temperature (RT), the MT process, and the chemical composition of the $Ni_{55}Fe_{16}Ga_{26}$ samples utilized in this study can be also found in our earlier publication [21].

Thermal properties of the samples were assessed using two distinct instruments: a commercial differential scanning calorimeter (DSC) Perkin-Elmer DSC7 (Perkin-Elmer, Norwalk, CT, USA) and a home-made conduction calorimeter able to perform high-resolution DTA analysis (1 nW) [26, 27].

DSC was employed under Ar flow and equipped with a cooling system. Various heating rates, $\beta$, (ranging from 5 to 40 K min$^{-1}$) were applied. To take into account the impact of the varying $\beta$ on the measured temperature, the melting temperature of In standard (429.75 K) was utilized (errors below 0.5 K) when the cooling system was active, and the melting point of Pb (600.65 K) was used otherwise.

For the ribbons, an initial step heating up to 473 K was applied to eliminate the influence of heterogeneities due to strain fields that might be present in the sample during its processing, thereby ensuring consistency and repetitiveness in subsequent measurements. Moreover, to mitigate potential variations in the characteristic parameters of the forward and reverse MT caused by cycling and inhomogeneities [24], the same ribbon pieces were consistently used for the same series of DSC measurements. Given the small mass of an individual piece relative to the crucible dimensions of the equipment, multiple pieces of ribbons were included in each experiment to achieve a mass comparable to that of the utilized standards. For sake of comparison, a consistent protocol was applied to the bulk sample, but using a single piece for all experiments.

In TME experiments using DSC, samples were initially heated at $\beta$=20 K min$^{-1}$ until $T_{stop} \in (A_s, A_f)$ was reached and the temperature was halted for 10 min. This time was chosen from our previous results [18] that show that isothermal treatments in these samples require 10 min of dwell time to stabilize the transformed phase fraction. Therefore, after 10 min, we expect that temperature will be homogeneous and athermal processes are completed. Then, the samples were cooled below $M_f$ at constant $\beta$=-20 K min$^{-1}$. In the subsequent step, the samples were reheated above $A_f$ at a constant rate, while employing varying rates from 5-40 K min$^{-1}$ for the second heating step to investigate the kinetics of the process. Additionally, different $T_{stop}$ values were utilized. To clarify the followed process, Fig. 1 illustrates the temperature-time curve of a complete DSC experiment, including an isothermal dwell conducted at 381 K, interrupted after 10 minutes, as an illustrative example. The inclusion of an isothermal step is necessary to achieve uniform temperature distribution within the crucible, particularly because multiple pieces of ribbons are present. The marked zone corresponds to that in which the heating rate has been modified for different experiments. The final temperature of second heating, which exceeds $A_f$, ensures the elimination of any residual memory effects, as the sample is fully transformed into the austenitic phase. Starting the heating process after stabilization of the calorimeter below $M_f$ confirms that the sample is initially in the martensitic state.

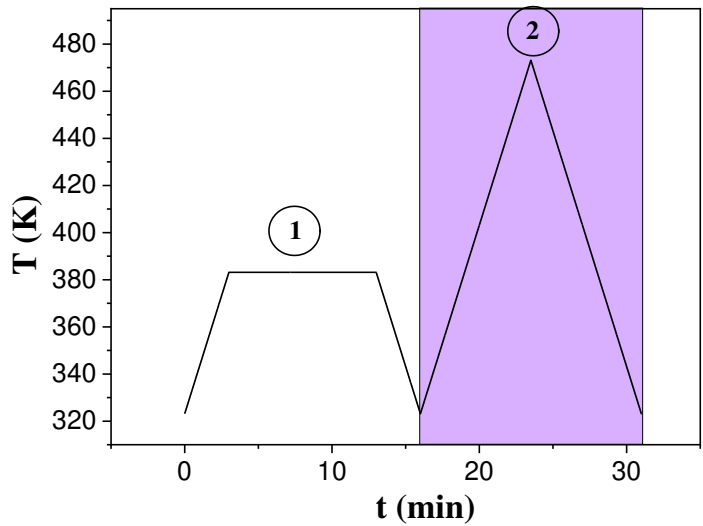

**Fig. 1.** *Illustration of a comprehensive thermal treatment conducted in commercial DSC experiments to examine the thermal memory effect in the investigated NiFeGa ribbon samples. In the initial heating step ($\beta$=20 K min$^{-1}$), the reverse martensitic transformation was halted at a specific temperature, 381 K and for a duration of 10 minutes in the reported case (1). Subsequently, the sample underwent cooling below martensite finish temperature and was then*

Thermal properties were also studied in the conduction calorimeter capable of capturing high-resolution DTA traces. Detailed information about the equipment and its functionality can be found in refs. [26, 27]. Due to the substantial thermal inertia of the equipment, scanning rates vary from a few kelvins per hour (slow rate range) to a few millikelvins per hour (ultraslow rate range), significantly slower than those used in commercial calorimeters. The sample was a parallelepiped piece with dimensions 9.0 mm x 9.57 mm x 2.35 mm (length x width x height) and 1.5987(1) g in mass.

Surface observations of the specimens were performed using a FEI Teneo scanning electron microscope (SEM) operating at 20 kV. Before SEM imaging, the surface of bulk and ribbon samples was mechanically polished using a series of sandpapers with varying grit sizes, followed by polishing with abrasive pastes.

## 3. RESULTS

In this study, we have analyzed six different samples, namely the as-spun ribbon (AS-R), thermally treated ribbon (TT-R), pressure treated ribbon (PT-R), pulverized ribbon (powder), as-prepared bulk (AP-B), and thermally treated bulk (TT-B). Table 1 presents the key transition temperatures, encompassing the austenite start, $A_s$, and finish, $A_f$, temperatures, along with the martensitic start, $M_s$, and finish, $M_f$, temperatures. Furthermore, the respective structure of each sample at RT determined by X-ray diffraction technique is outlined.

The structure of the AS-R sample displays a monophasic martensite with a modulated 14M structure at RT. However, when the ribbon is crushed into powder, an intermartensitic transformation occurs, leading to a non-modulated L1$_0$ structure [20]. When uniaxial pressure is applied on the ribbons, a partially intermartensitic transformation between the 14M modulated to a non-modulated phase occurs [20]. On the other hand, thermal treatments can induce the precipitation of the $\gamma$ phase [21]. Conversely, the arc-melting method fails to yield a single-phase sample, resulting in the simultaneous presence of a non-modulated martensite structure and $\gamma$ phase at RT. In the bulk sample, the modulated phase emerges following an extended heat treatment (1073 K for 24 h) and subsequent rapid water quenching, with a small increase of $\gamma$ phase after this treatment. The characteristic temperatures of the MT are lower in the bulk sample compared to the ribbon, influenced by the alteration of the $e/a$ parameter due to the presence of the $\gamma$ phase with a lower content in Ga [21].

Table 1. Characteristic temperatures associated with the MT measured by DSC scans at 20 K min$^{-1}$ and the structure of the analyzed samples at RT. ND: Non-detected; AS: As-spun; AP: As-prepared; TT: Thermally-treated; PT: Pressured-treated. In the case of the PT sample, the presented data correspond to the MT of the 14M structure.

| Sample | | Structure | $M_s$ (K) | $M_f$ (K) | $A_s$ (K) | $A_f$ (K) |
|---|---|---|---|---|---|---|
| **Ribbon** | AS | 14M | 370.3 | 346.9 | 379.5 | 399.5 |
| | TT | 14M+$\gamma$ | 360.6 | 338.5 | 371.2 | 388.0 |
| | PT | 14M+L1$_0$ | 370.3 | 346.9 | 379.5 | 399.5 |
| | Powder | L1$_0$ | ND | ND | ND | ND |
| **Bulk** | AP | L1$_0$+$\gamma$ | 321.1 | 288.4 | 322.9 | 353.0 |
| | TT | 14M+$\gamma$ | 300.3 | 283.3 | 311.9 | 324.8 |

The microstructure of the samples was analyzed using SEM. Figure 2 shows representative backscattered electron (BSE) images of the AS ribbon (panels a and b), TT ribbon (panels c and d), and the powder derived from the ribbons (panels e and f). Similarly, Figure 3 shows BSE images of the AP bulk sample (panels a and b) and the TT bulk sample (panels c and d).

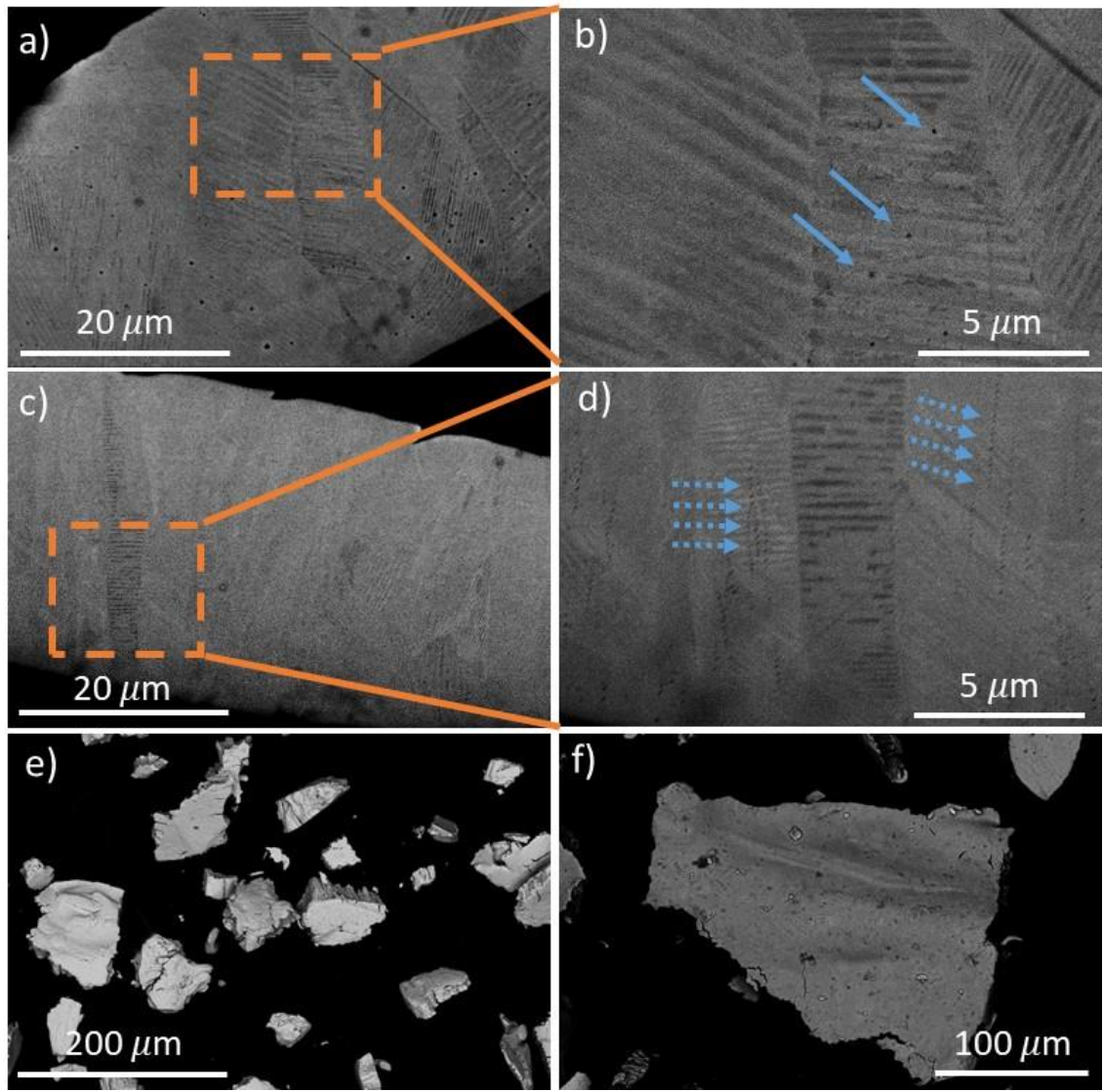

**Figure 2.** *SEM micrographs using backscattered electrons at different magnifications: a) and b) as-spun ribbon, c) and d) thermally treated ribbon, and e) and f) powder obtained from manually grinding the as-spun ribbon. Figures b) and d) shows higher magnifications of selected areas of panels a) and c), respectively.*

The thickness of the ribbons varied slightly, ranging from 30 to 40 $\mu$m (Fig. 2a), and predominantly featured a single-phase martensitic microstructure. This microstructure consisted of several variants of martensite plates with coarse grains measuring several microns in size. The martensitic lamellas, exhibiting a plate-like shape, were clearly visible, along with small black pores (indicated with arrows) formed during the quenching process from the melt. No significant differences were observed between the free and wheel side of the ribbons. The microstructure shows few grain boundaries, in which grain growth was inhibited by the finite thickness of the ribbons [28]. Within the martensite plates, conjugation boundaries, CB, were visible, separating regions with differently oriented nanotwins [29]. These CBs were commonly found in the self-accommodated state of martensite microstructure, particularly near grain boundaries where branching occurs [30]. Branching is realized by a shift in nanotwin orientation to its conjugated counterpart within a single martensitic plate, forming CBs [31].

In the TT-R, the thickness of the martensite plates increased, even occupying the entire cross-section of the ribbon. Interestingly, despite the relatively low treatment temperature and the short annealing time, this process appeared to trigger the precipitation of an additional phase dispersed within the matrix (indicated by dashed arrows in Fig. 2d). The precipitated phase was discernible by its dark grey contrast and strong alignment roughly perpendicular to the surface of the ribbon. Figures 2d and 2f display SEM-BSE images of powders produced by manually milling the melt-spun ribbons. The particle sizes is not homogeneous and ranged from 50 to 150 $\mu$m, predominantly exhibiting a polygonal shape, which resulted from the brittleness of ribbons due to the absence of gamma precipitates.

Figure 3 confirms the significant variation in microstructure depending on the preparation method. Specifically, while the AS-R samples displayed a single-phase microstructure, the samples produced using the arc-melting technique revealed the presence of a clearly distinguishable second phase. The morphology of this second phase changed with thermal treatment, leading to an increase in the size of the precipitates. The precipitation of the gamma phase, which is enriched in Fe and impoverished in Ga, reduced the Fe content in the martensite matrix. This is qualitatively shown in [25] by EDS measurements. Although the quantitative changes are too small, qualitative behavior is clear and may lead to an increase in electrons per atom (e/a) factor, thereby lowering the martensitic transformation temperatures. It is also worth noting that the distribution of precipitates in Fig. 2d follows the same patterns as those observed in Fig. 3a.

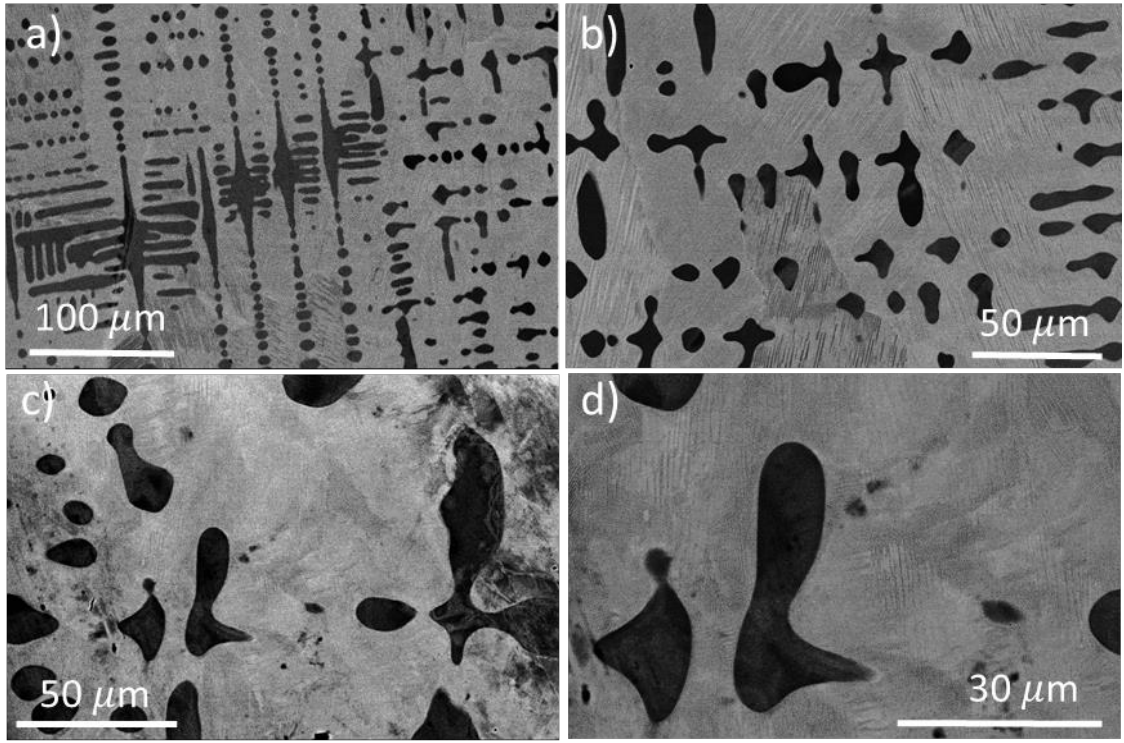

**Figure 3.** *SEM micrographs using backscattered electrons: a) and b) as-prepared bulk, and c) and d) thermally treated bulk at different magnifications.*

Unlike ribbon samples, porosity is not observed in bulks. Moreover, the low temperature at which the ribbons have been treated are not expected to produce any evolution in the porous density. Therefore, although this difference may also influence the varying behavior observed between ribbon and bulk samples, it is unlikely to account for the differences observed specifically among ribbon samples.

**3.1. Temperature Memory Effect in $Ni_{55}Fe_{19}Ga_{26}$ as-melt spun ribbons**

Figure 4 shows (dashed lines) the direct and reverse transformation in the AS-R samples after conventional heating (reverse MT) and cooling runs (direct MT). It is important to emphasize that the dashed line, representing the entire thermal cycle of a relaxed sample, is consistently featured in the DSC plots for comparative purposes.

We choose $\beta = \pm 20$ K min$^{-1}$ to optimize the signal-to-noise ratio in DSC measurements [18]. Notably, the ribbon specimens underwent two pre-thermal cycles from RT to 473 K (relaxed sample) to remove potential inhomogeneities among different ribbon pieces, influenced by previous thermo-mechanical histories [21]. The actual procedure for testing the TME involved: i) Heating from RT to $T_{stop}$ ($A_S$<$T_{Stop}$<$A_f$); ii) Kinetic stop at $T_{stop}$ for 10 min; iii) Cooling down to 323 K; and iv) Heating from 323 K to 473 K. Steps i) and ii) could be repeated multiple times with varying $T_{stop}$ in each incomplete reverse MT.

Figures 4a and 4b display the characteristic curves obtained from a TME test with a single stop, where the $T_{stop}$ was set at 381 K and 383 K, respectively. Notably, kinetic stops are clearly discernible on the heat flow curves during the heating. In addition to the single-stop experiments, a double-stop test was executed. Figure 2c displays DSC scans after performing two consecutive incomplete cycles during the heating phase at temperatures of 383 K and 381 K in decreasing order. In such case, two kinetic stops show up in the subsequent complete reverse MT. These findings are consistent with previous results, that associate the number of thermal arrests arranged in decreasing $T_{stop}$ with the number of interruptions in the subsequent heating run. In contrast, when $T_{stop}$ is arranged in an increasing order, then one kinetic stop at the highest temperature $T_{stop}$ is only observed [4, 32, 33].

Incomplete direct MT were also tested on cooling experiments, incorporating different $T_{stop}$ (between $M_f$ and $M_s$) and durations. Here, only a fraction of the austenite undergoes transformation into the martensite phase, leaving the rest of the austenite unaffected. After this interrupted direct MT, the transformed martensite has the potential to revert to austenite upon further temperature elevation beyond $A_f$. However, the outcomes reveal that the subsequent cooling scans registering direct MT does not display any indications of kinetic interruption. These results are in line with those previously reported [4, 16].

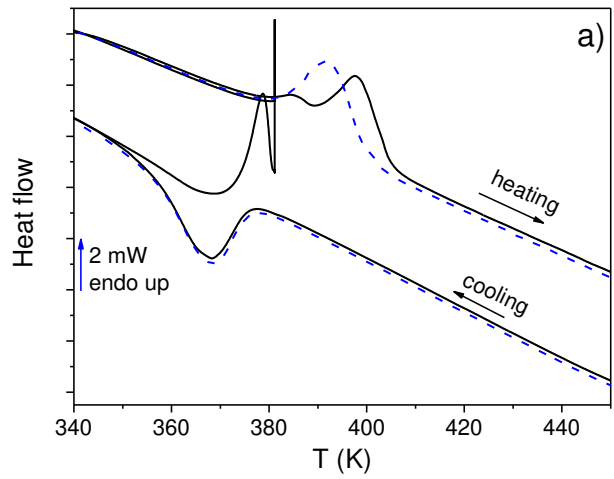

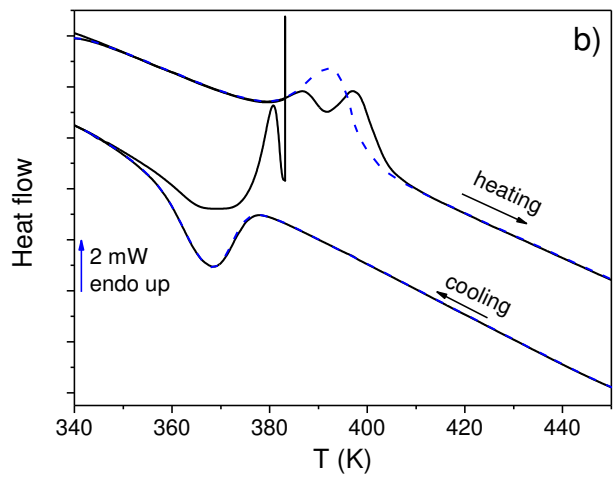

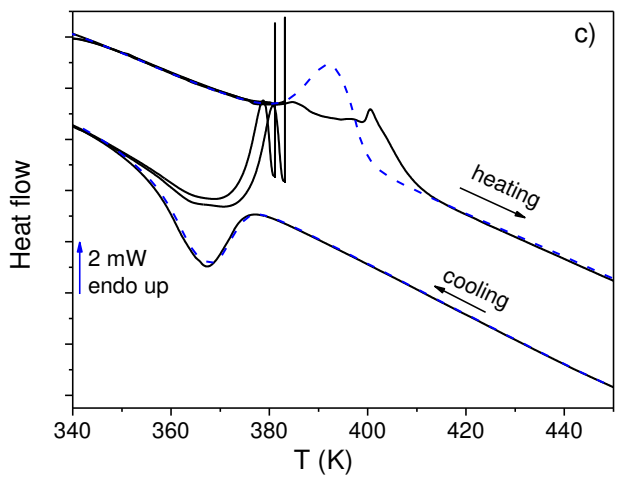

***Figure 4.** DSC scans registered using the protocol of figure 1 (continuous black lines) along with complete DSC scans (dashed blue lines) representing the transition from reverse and direct MT in $Ni_{55}Fe_{19}Ga_{26}$ ribbons. These scans were performed following either a singular arrest at $T_{stop1}=381$ K (shown in panel a), a singular arrest at $T_{stop2}=383$ K (illustrated in panel b), and a combination of arrest at $T_{stop2}$ and $T_{stop1}$, each lasting 10 minutes at their respective temperatures (depicted in panel c).*

## 3.2. Temperature Memory Effect in Ni$_{55}$Fe$_{19}$Ga$_{26}$ thermally treated ribbons

Previous studies show that heat treatments conducted above the MT led to an irreversible downward shift in MT due to structural modifications. In the case of thermal treatments performed below 573 K, stress relaxation accumulated during ribbon fabrication occurred. Annealing treatments below 873 K cause slight changes in the lattice parameters, altering the $b/a$ and destabilizing the 14M modulated structure. At temperatures above 873 K, the monophasic nature of the ribbon is lost, and the martenstic modulated structure transforms into two distinct phases: austenite and the $\gamma$-phase [21]. In order to analyze the differences in TME due to these microstructural changes, TT-R samples were obtained after heating AS-R samples up to 623 K.

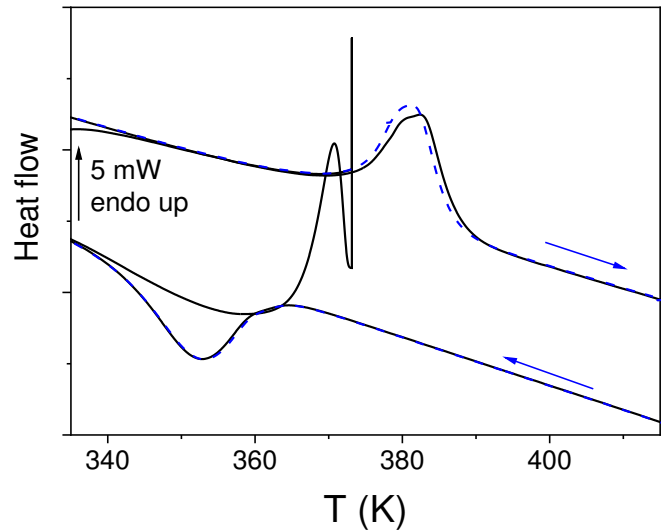

*Figure 5.* DSC scan registered using the protocol of figure 1 using a $T_{stop}$=373 K and a dwell time of 10 min (continuous black line) along with complete DSC scan (dashed blue line) capturing the transition from martensite to austenite in the Ni$_{55}$Fe$_{19}$Ga$_{26}$ ribbon after heated up to 623 K (TT-R sample).

Figure 5 illustrates the DSC scans for the Ni$_{55}$Fe$_{19}$Ga$_{26}$ ribbon after being heated up to 623 K. $M_f$ and $M_s$ have decreased with the thermal treatment in comparison with the AS-R. The sample was heated until the MT reverse transformation was halted at 373 K for 10 min. Then, the sample was cooled down to 323 K and the subsequent heating run showed a smoother kinetic interruption. This evidence suggests that the destabilization of the modulated structure ($b/a$ <7) was smearing the TME. Increasing isothermal dwell time up to 60 min does not enhance the TME.

The phase fraction of the 14M structure in the TT-R sample can be inferred from the enthalpy change observed in the TT-R Sample and the AS-R sample. Assuming that the AS-R is 100% in 14M ($\Delta H$=5.46 J g$^{-1}$), then we infer ~90% of TT-R is 14M phase ($\Delta H$=5.06 J g$^{-1}$).

### 3.3. Temperature memory effect in Ni$_{55}$Fe$_{19}$Ga$_{26}$ mechanically treated ribbons

The application of a uniaxial pressure on the AS-R sample induces a gradual transformation from the 14M modulated phase to the non-modulated L1$_0$ structure [20]. Specifically, a load of 2 tons was applied to a 0.785 cm$^2$ surface (~250MPa) for 5 minutes at RT to produce PT-R sample.

DSC scans of PT-R sample are depicted in Figure 6. The distinct peaks attributed to the MT observed in the AS-R exhibit a noticeable decrease. This suggests that the 14M, the transformable phase, becomes destabilized under pressure. From $\Delta H$=2.64 J g$^{-1}$, we infer that ~50% of the PT-R sample is 14M, the remaining half belongs to the non-modulated L1$_0$ phase, as the $\gamma$ − phase is not induced by pressure [20]. Additionally, the interrupted reverse MT conducted at the same $T_{stop}$ than those employed in Fig. 1 reveals some TME.

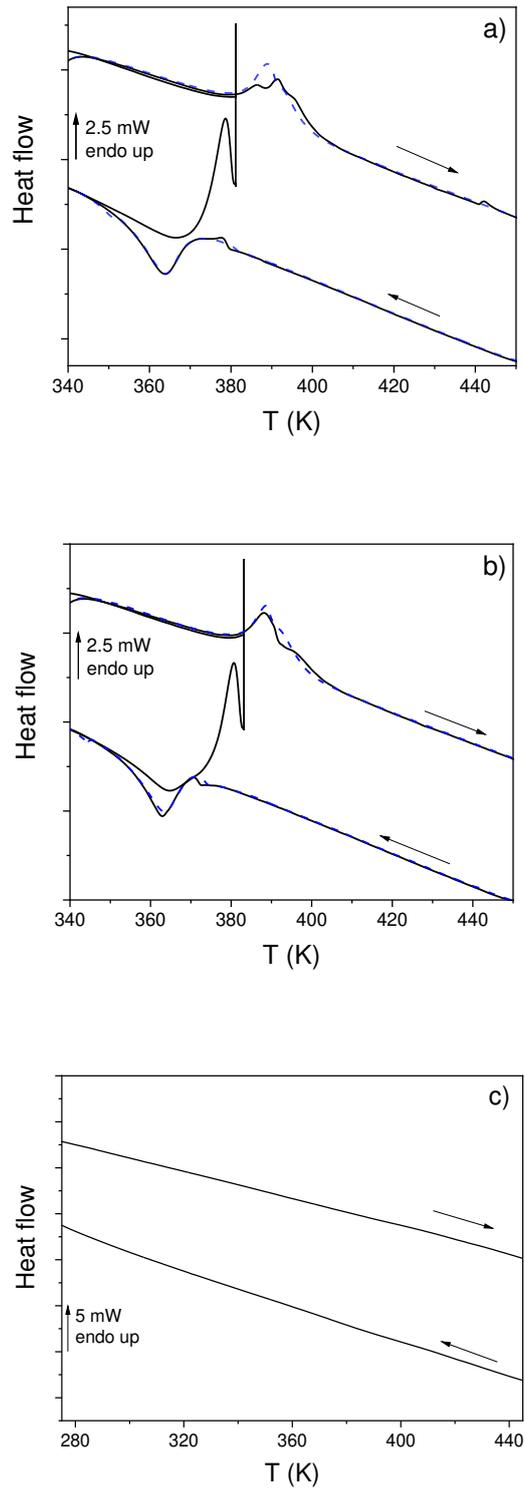

***Figure 6.*** *DSC scans registered using the protocol of figure 1 (continuous black line) with $T_{stop}$ set at a) 381 K and b) 383 K, along with complete DSC scan (dashed blue line) capturing the martensitic transformation in the $Ni_{55}Fe_{19}Ga_{26}$ PT-R samples. Panel c) corresponds to DSC heating and cooling curves of the $Ni_{55}Fe_{19}Ga_{26}$ powders derived from as-prepared ribbons.*

Figure 6c depicts the DSC plots of the $Ni_{55}Fe_{19}Ga_{26}$ powders derived from as-prepared ribbons, exhibiting a $L1_0$ monophasic martensite structure (see ref. [20]). Under the tested conditions, no MT is observed, indicating the elimination of the modulated phase through mechanical treatment. These findings align with prior research on Ni-Mn-Sn milled alloys, where the peak associated with MT diminishes notably with prolonged milling time, eventually disappearing after 45 minutes [34].

### 3.4. Temperature memory effect in $Ni_{55}Fe_{19}Ga_{26}$ as-prepared bulk

Figure 7 displays the DSC scans of the $Ni_{55}Fe_{19}Ga_{26}$ AP-B sample, featuring a biphasic structure ($L1_0 + \gamma$ precipitates [21]), following an interrupted reverse MT at a $T_{stop}$= 323 K, between $A_s$ and $A_f$ during heating. In all tested conditions, there is no indication of a kinetic interruption for the subsequent complete transformation. This apparent absence of TME aligns with those documented in [17], where bulk NiFeGa samples did not exhibit distinct dips linked to TME after an incomplete reverse phase transition.

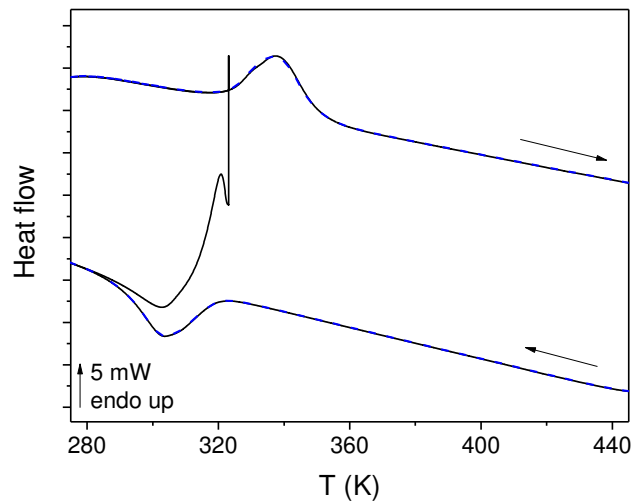

*Figure 7. DSC scans registered using the protocol of figure 1 (continuous black line) with $T_{stop}$=323 K and dwell time set at 10 min, along with complete DSC scan (dashed blue line) capturing the transition from martensite to austenite in the $Ni_{55}Fe_{19}Ga_{26}$ as-prepared bulk.*

## 3.5. Temperature Memory Effect in thermally-treated bulk

The NiFeGa TT-B sample exhibits a biphasic structure with 14M martensite phase and $\gamma-$ phase precipitates. Figure 8 shows the DSC results. With a $\Delta H=$ 3.30 J g$^{-1}$ and from the previous $\Delta H$ for the AS-R sample we infer a $\sim$60% 14M and $\sim$40% $\gamma-$ phase mixture.

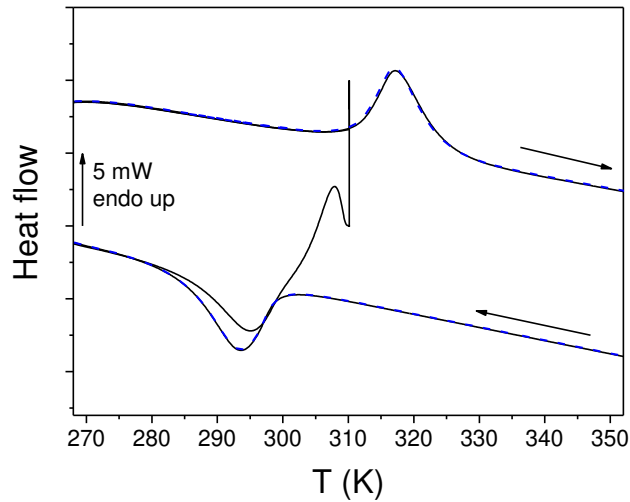

*Figure 8. DSC scans registered using the protocol of figure 1 (continuous black line) with $T_{stop}$=310 K and dwell time set at 10 min, along with complete DSC scan (dashed blue line) capturing the transition from martensite to austenite in the Ni$_{55}$Fe$_{19}$Ga$_{26}$ thermally treated bulk.*

The TT-B sample was also placed in a home-made conduction calorimeter that operates as a high-sensitive DTA tracer. The sample sustained two ultraslow heating runs. As a reference, it was heated at a rate of 40 mK h$^{-1}$ and the reverse transformation was recorded. Results from this experiment were shown previously in ref. [25]. Then, the sample was cooled from the austenite phase and heated again at a rate of 40 mK h$^{-1}$. The temperature was halted for five days at 315.7 K. The sample was then cooled down at a rate of 40 mK h$^{-1}$ until 309 K was reached. It was then quenched down to 280 K, heated back to 306 K when the standard heating profile was resumed, and the sample completed the reverse MT. The heating DTA traces in the temperature region close to the MT are shown in

Figure 9. A kinetic stop is clearly discernible on the heat flow curves during the heating after a smooth baseline was subtracted.

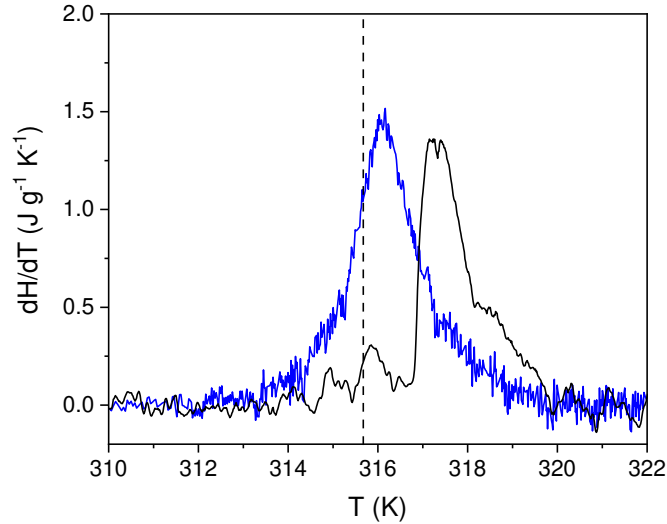

*Figure 9.* DTA traces for TT-B sample without thermal arrest (reference in blue) and after thermal arrest at 315.7 K (black).in the temperature ranges around reverse martensitic transformation.

## 4. DISCUSSION

MT temperatures can vary within a certain temperature range, depending on $\beta$ [19, 35]. Therefore, calorimetric measurements after thermal arrest were conducted with $\beta$ ranging from 5 to 40 K min$^{-1}$ to identify the dependence of the MT temperature interval after kinetic arrest. Initially, heating to $T_{stop}$ followed by subsequent cooling to 323 K was conducted at 20 K min$^{-1}$. Subsequently, a heating scan up to 473 K was performed at different $\beta$ ranging from 5 to 40 K min$^{-1}$.

The resulting DSC scans are illustrated in Figure 10 (left panels). Panels a and b depict the characteristic curves obtained from a single $T_{stop}$ of 381 K and 383 K, respectively. Additionally, a double test was conducted, as shown in panel c. A slight shift towards higher temperatures is observed for $A_f$ as $\beta$ increases, while $\beta$ has no discernible influence on $A_s$. Consequently, the transformation temperature interval $\Delta T = A_f - A_s$ increases with $\beta$. These results are consistent with those reported by Wang et al., suggesting that $\beta$ leads to a larger variation in $A_f$ compared to $A_s$ [36].

Right panels of Fig. 10 show the corresponding transformed fractions which were approximated by the normalized transformed enthalpy, $X = \Delta H(T)/\Delta H_{total}$, where $\Delta H(T)$ represents the enthalpy developed up to temperature $T$, and $\Delta H_{total}$ denotes the total enthalpy of the MT. All curves collapse into a common line below $T_{stop}$, within the standard uncertainty associated to baseline determination in the DSC [37]. However, deviations can be observed above $T_{stop}$ in agreement with the $\beta$ effect described above. When comparing the obtained curves with those of a complete transformation without any interruption, a kinetic delay evidenced by a shift of the MT to higher temperatures (~5K) from $T_{stop}$ is observed. This delay seems to be accumulative as observed in the experiment with two $T_{stop}$ in Figure 10 f, with a delay >10 K.

The impact of the TME on the evolution of the transformed fraction in the other studied samples has also been analyzed and depicted in Figure 11. As anticipated, the shift of $X$ in the case of the TT-R (Fig. 11a) and PT-R (Fig. 11b) samples is significantly reduced compared to the AS-R, with a shift to higher temperatures of approximately 1 K. However, in the case of the bulk samples, this shift is within the resolution of the equipment, regardless of the martensite phase structure (see panels c and d for the AP-B and TT-B samples, respectively).

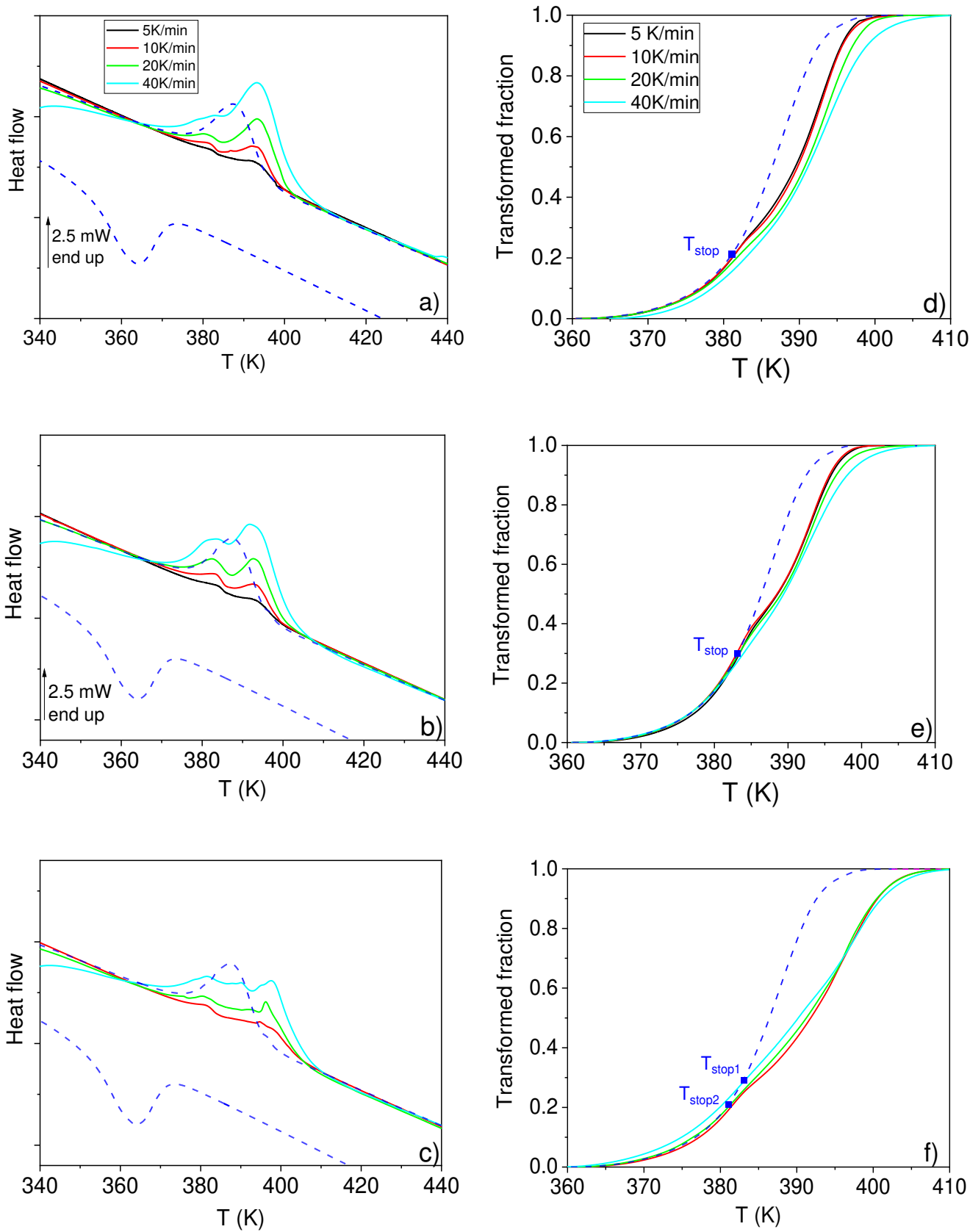

*Figure 10*. *DSC scans registered (a,b,c) and corresponding transformed fractions (d,e,f) using the protocol of figure 1 (continuous lines) along with DSC signal for untreated sample (dashed blue lines) representing the transition from reverse and direct MT in $Ni_{55}Fe_{19}Ga_{26}$ ribbons at the indicated heating rates. These scans were performed following either a singular arrest at $T_{stop1}=381$ K (a,d), a singular arrest at $T_{stop2}=383$ K (b,e), and a combination of arrest at $T_{stop2}$ and $T_{stop1}$, each lasting 10 minutes at their respective temperatures (c,f).*

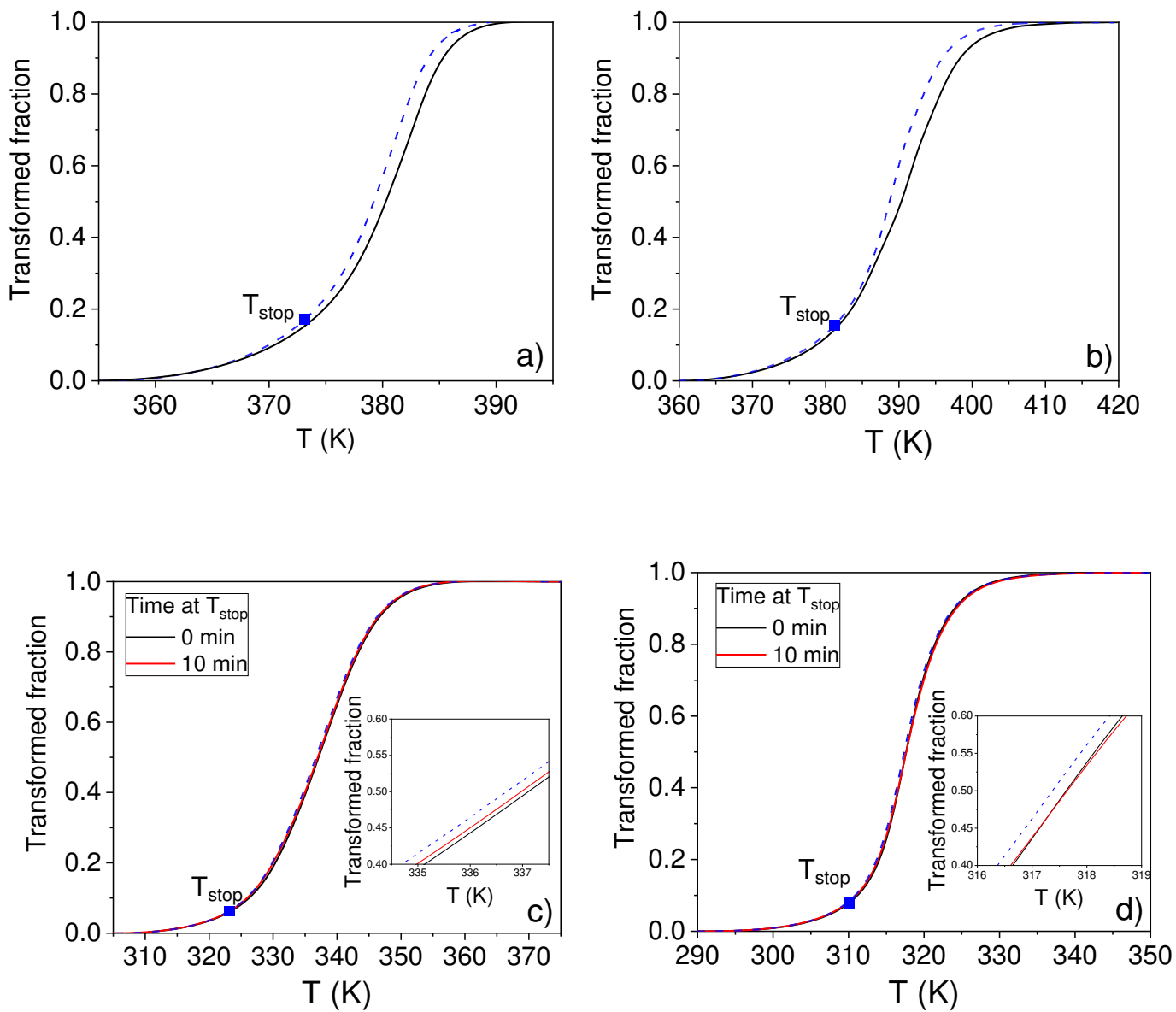

**Figure 11**. *Transformed fraction as a function of temperature of the reverse MT after kinetic arrest in the case of the a) TT-R and b) PT-R samples; and c) AP-B and d) TT-B samples. Insets show the enlarged 0.4 < X < 0.6 region.*

Figure 12 shows the results for the high-resolution DTA experiment corresponding to the TT-B sample. In ultraslow DTA experiment, the transformed fraction was approximated by the normalized transformed enthalpy from 310 K (where the austenite fraction is zero) and assuming that at 323 K, the austenite phase fraction is 1. The reduction in the width of reverse MT is in agreement with the heating rate effect described above. When comparing the transformed fraction with those of the complete reverse MT without interruption (dashed-line in Fig.12), a slight shift to higher temperatures can be observed. Therefore, its behavior is similar to that shown by ribbon samples after analyzing TME (see Figure 10), but not in the case of the bulk. The higher resolution of the DTA scans at very low heating rate, along with the very long isotherm dwell, enables us to assess that TME also occurs in the TT-B sample.

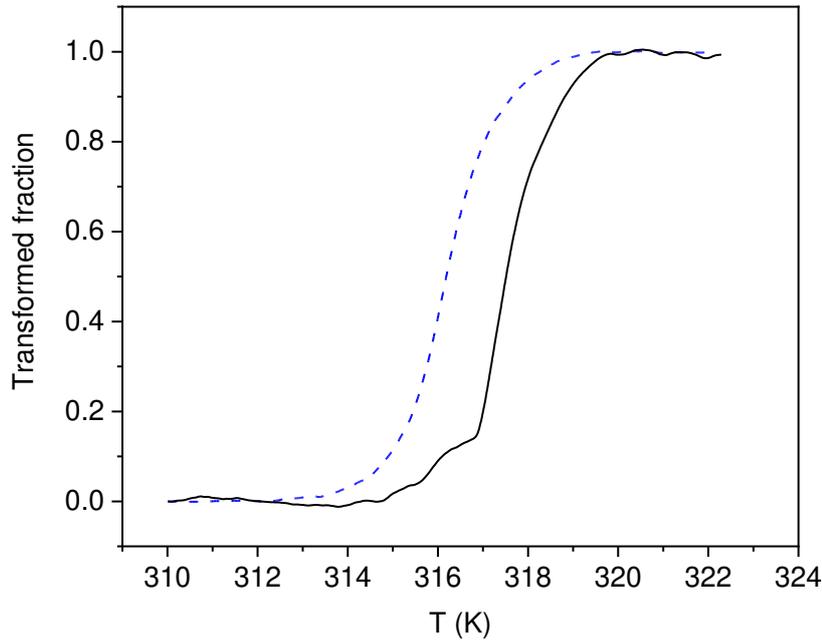

**Figure 12**. *Transformed fraction as a function of temperature of the reverse MT after kinetic arrest in the TTB sample in the ultraslow calorimeter along as of the original transformed fraction (dashed line) for original scans.*

The TME observed in ultraslow DTA for TT-B sample is higher than that presumed for DSC experiments. This difference could be due to the very long isotherm and the effect of reduced $\beta$, which could be more efficient in releasing strain in the untransformed martensite phase, as it will be discussed below, to be a suitable mechanism for TME.

In TME experiments, during the partial reverse MT, the transition from martensite to austenite halts at a specific temperature between $A_s$ and $A_f$. In that situation, only a portion of the martensite reverts to the austenite phase, while the rest of the martensite persists. This residual martensite is commonly referred as M1. Upon subsequent cooling down to $M_f$, the austenite phase undergoes re-transformation into martensite, resulting in the formation of a new martensite phase, denoted as M2. During the next heating scan, M2 and M1 sequentially transit to the austenite phase, leading to a kinetic delay between both transformations.

Several studies have tried to explain the TME by attributing it to the release of elastic strain energy during the first interrupted transformation in the remnant M1. This stored elastic strain would act as instabilities to launch nucleation process in the reverse transformation [2]. Therefore, M1 is stabilized in a second reverse transition as the elastic strain is lowered. However, the proposed mechanisms considerably vary and are subject to debate. Madangopal et al. [8] suggested that M2 might accumulate more elastic strain energy than M1, thereby potentially pre-positioning the reverse transformation of M2. However, we do observe a common $A_s$ value independently of the presence of TME (or even a delay when very long isotherms occur at $T_{stop}$) but a delay that initiate approximately at the temperature where the halt have occurred (in the absence of very long isotherm at $T_{stop}$). Conversely, it has been proposed that the TME arises from the reduction of elastic strain energy in M1, requiring a higher temperature to initiate the transformation of M1 once the transformation of M2 is complete [9]. Our results are in agreement with this second interpretation.

Upon cooling, the atomic arrangement adopts a twinning structure with periodic stacking order, potentially resulting in 24 variants of martensite [13]. The MT stores some elastic strain energy within the thermoelastic martensite variants. Therefore, while the release of elastic strain energy from martensite variants is often associated with the TME, it is important to note that the coherent energy of adjacent phases also influences the TME [38]. This effect is clearly observed in our results. On the one hand, the presence of the non-modulated phase and the destabilization of the 14M phase restricts the occurrence of TME (see Figs. 5 and 6).Conversely, in bulk samples, the presence of around 40% $\gamma$ precipitates inhibits TME to require high precision techniques to evidence it.

## 5. CONCLUSIONS

The study of the thermal memory effect, TME, involved six samples sharing the $Ni_{55}Fe_{19}Ga_{26}$ composition but differing in micro and macrostructure. The results show that the TME, triggered by a partial reverse transformation from martensite to austenite phase, is not a common phenomenon in the studied shape memory alloys. Instead, it appears to be contingent upon the specific microstructure of the sample. Remarkably, this effect was clearly detected by DSC in the as-spun ribbon sample (kinetic delay about 5 K) with single phase modulated 14M martensite structure at room temperature. Various heating rates were applied to confirm the athermal character of the reverse martensitic transformation even after TME. The temperature interval in which the transformation occurs increases with the heating rate.

Thermal treatment in ribbon samples leading to destabilization of modulated 14M and production of $\gamma -$ phase reduces TME (kinetic delay <5 K) as it does pressure application leading to stabilization of non-modulated martensite. Conversely, in bulk samples, which are not monophasic and present $\gamma$ precipitates (60 % of the 14M phase is estimated in the case of the thermally treated bulk), TME is not detected by DSC (kinetic delay <0.2 K). However, quasistatic DTA experiments evidenced that TME exists in thermally treated bulk sample. Additionally, powdering the as-melt spun ribbon sample yields the disappearance of the MT signal in the DSC experiments conducted in this study.

No thermal arrest phenomena were observed in direct martensitic transition during cooling. The obtained results indicate the important effect of the adjacent phases on the TME; the destabilization of 14M phase and $\gamma$ precipitates clearly hinder the occurrence of the phenomenon.

**Acknowledgements**

This research was funded by Junta de Andalucía-Consejería de Conocimiento, Investigación y Universidad, under project with reference ProyExcel_00360 and supported by the PAI of the Regional Government of Andalucía and the VI and VII-PPITU from Universidad de Sevilla (Spain). A. Vidal-Crespo acknowledges the financial support of the VI-PPITU from Universidad de Sevilla (Spain).

**Declaration of Competing Interest**

The authors declare that they have no known competing financial interests or personal relationships that could have appeared to influence the work reported in this paper.

**Declaration of Generative AI and AI-assisted technologies in the writing process**

During the preparation of this work the authors used ChatGPT-OpenAI in order to improve the language. After using this tool, the authors reviewed and edited the content as needed and take full responsibility for the content of the publication

**CRediT authorship contribution statement**

# References

[1] W.M. Huang, Z. Ding, C.C. Wang, J. Wei, Y. Zhao, H. Purnawali, Shape memory materials, Materials Today 13(7) (2010) 54-61.
[2] J.W. Christian, The theory of transformations in metals and alloys. I, Equilibrium and general kinetic theory 586 (1975).
[3] Y. Zheng, L. Cui, J. Schrooten, Temperature memory effect of a nickel–titanium shape memory alloy, Applied Physics Letters 84(1) (2004) 31-33.
[4] Z.G. Wang, X.T. Zu, Y.Q. Fu, L.M. Wang, Temperature memory effect in TiNi-based shape memory alloys, Thermochimica Acta 428(1) (2005) 199-205.
[5] J. Rodríguez-Aseguinolaza, I. Ruiz-Larrea, M.L. Nó, A. López-Echarri, J.S. Juan, A new quantitative approach to the thermoelastic martensitic transformation: The density of elastic states, Acta Materialia 56(20) (2008) 6283-6290.
[6] J. Rodríguez-Aseguinolaza, I. Ruiz-Larrea, M.L. Nó, A. López-Echarri, J. San Juan, The influence of partial cycling on the martensitic transformation kinetics in shape memory alloys, Intermetallics 17(9) (2009) 749-752.
[7] F. Ţolea, M. Ţolea, M. Sofronie, M. Văleanu, Distribution of plates' sizes tell the thermal history in a simulated martensitic-like phase transition, Solid State Communications 213-214 (2015) 37-41.
[8] K. Madangopal, S. Banerjee, S. Lele, Thermal arrest memory effect, Acta Metallurgica et Materialia 42(6) (1994) 1875-1885.
[9] G. Airoldi, A. Corsi, G. Riva, Step-wise martensite to austenite reversible transformation stimulated by temperature or stress: a comparison in NiTi alloys, Materials Science and Engineering: A 241(1) (1998) 233-240.
[10] G. Airoldi, A. Corsi, G. Riva, Step-wise martensite to austenite reversible transformation stimulated by temperature or stress: a comparison in NiTi alloys, Materials Science and Engineering: A 241(1-2) (1998) 233-240.
[11] X. He, J. Xiang, M. Li, S. Duo, S. Guo, R. Zhang, L. Rong, Temperature memory effect induced by incomplete transformation in TiNi-based shape memory alloy, Journal of alloys and compounds 422(1-2) (2006) 338-341.
[12] J. Rodriguez-Aseguinolaza, I. Ruiz-Larrea, M. Nó, A. Lopez-Echarri, J. San Juan, Thermodynamic study of the temperature memory effects in Cu–Al–Ni shape memory alloys, Journal of Applied Physics 107(8) (2010).
[13] K. Otsuka, C.M. Wayman, Shape memory materials, Cambridge university press1999.
[14] I. Ruiz-Larrea, A. López-Echarri, J.F. Gómez-Cortés, M.L. Nó, D.W. Brown, L. Balogh, T. Breczewski, J. San Juan, Strain relaxation in Cu-Al-Ni shape memory alloys studied by in situ neutron diffraction experiments, Journal of Applied Physics 125(8) (2019).
[15] M.Z. Zhou, X. Zhang, X.L. Meng, W. Cai, L.C. Zhao, Temperature Memory Effect Induced by Incomplete Transformation in Ni-Mn-Ga-based Shape Memory Alloy, Materials Today: Proceedings 2 (2015) S867-S870.
[16] M.Z. Zhou, W.M. Huang, X.L. Meng, Temperature memory effect in a magnetic shape memory alloy for monitoring of minor over-cooling, Scripta Materialia 127 (2017) 41-44.
[17] F. Ţolea, M. Ţolea, M. Văleanu, Thermal memory fading by heating to a lower temperature: Experimental data on polycrystalline NiFeGa ribbons and 2D statistical model predictions, Solid State Communications 257 (2017) 36-41.
[18] A. Vidal-Crespo, A.F. Manchón-Gordón, J.S. Blázquez, J.J. Ipus, P. Svec, C.F. Conde, Thermal arrest analysis of the reverse martensitic transformation in a Ni55Fe19Ga26 Heusler alloy obtained by melt-spinning, Journal of Thermal Analysis and Calorimetry 148(6) (2023) 2367-2375.
[19] F. Tolea, M. Sofronie, A.D. Crisan, M. Enculescu, V. Kuncser, M. Valeanu, Effect of thermal treatments on the structural and magnetic transitions in melt-spun Ni-Fe-Ga-(Co) ribbons, Journal of Alloys and Compounds 650 (2015) 664-670.

[20] A.F. Manchón-Gordón, J.J. Ipus, M. Kowalczyk, A. Wójcik, J.S. Blázquez, C.F. Conde, W. Maziarz, P. Švec Sr, T. Kulik, A. Conde, Effect of pressure on the phase stability and magnetostructural transitions in nickel-rich NiFeGa ribbons, Journal of Alloys and Compounds 844 (2020) 156092.

[21] A.F. Manchón-Gordón, J.J. Ipus, M. Kowalczyk, J.S. Blázquez, C.F. Conde, P. Švec, T. Kulik, A. Conde, Comparative study of structural and magnetic properties of ribbon and bulk Ni55Fe19Ga26 Heusler alloy, Journal of Alloys and Compounds 889 (2021) 161819.

[22] Y. Boonyongmaneerat, M. Chmielus, D.C. Dunand, P. Müllner, Increasing Magnetoplasticity in Polycrystalline Ni-Mn-Ga by Reducing Internal Constraints through Porosity, Physical Review Letters 99(24) (2007) 247201.

[23] C. Witherspoon, P. Zheng, M. Chmielus, D.C. Dunand, P. Müllner, Effect of porosity on the magneto-mechanical behavior of polycrystalline magnetic shape-memory Ni–Mn–Ga foams, Acta Materialia 92 (2015) 64-71.

[24] A.F. Manchón-Gordón, R. López-Martín, J.J. Ipus, J.S. Blázquez, P. Svec, C.F. Conde, A. Conde, Kinetic Analysis of the Transformation from 14M Martensite to L21 Austenite in Ni-Fe-Ga Melt Spun Ribbons, Metals 11(6) (2021) 849.

[25] J.-M. Martín-Olalla, A. Vidal-Crespo, F.J. Romero, A.F. Manchón-Gordón, J.J. Ipus, J.S. Blázquez, M.C. Gallardo, C.F. Conde, Ultraslow calorimetric studies of the martensitic transformation of NiFeGa alloys: detection and analysis of avalanche phenomena, Journal of Thermal Analysis and Calorimetry  (2024).

[26] J.d. Cerro, S. Ramos, J.M. Sanchez-Laulhe, Flux calorimeter for measuring thermophysical properties of solids: study of TGS, Journal of Physics E: Scientific Instruments 20(6) (1987) 612.

[27] F. Jiménez, S. Ramos, J. Del Cerro, Specific heat measurement of LATGS crystals under nonequilibrium conditions, Phase Transitions 12(4) (1988) 275-284.

[28] P. Zheng, N.J. Kucza, C.L. Patrick, P. Müllner, D.C. Dunand, Mechanical and magnetic behavior of oligocrystalline Ni–Mn–Ga microwires, Journal of Alloys and Compounds 624 (2015) 226-233.

[29] B. Muntifering, R. Pond, L. Kovarik, N. Browning, P. Müllner, Intra-variant substructure in Ni–Mn–Ga martensite: conjugation boundaries, Acta materialia 71 (2014) 255-263.

[30] M. Szczerba, R. Chulist, Detwinning of a non-modulated Ni–Mn–Ga martensite: from self-accommodated microstructure to single crystal, Acta Materialia 85 (2015) 67-73.

[31] A. Brzoza, M. Kowalczyk, A. Wierzbicka-Miernik, P. Czaja, W. Maziarz, A. Wójcik, J. Wojewoda-Budka, M. Sikora, J. Dutkiewicz, M.J. Szczerba, Microstructural anisotropy, phase composition and magnetic properties of as-cast and annealed Ni-Mn-Ga-Co-Cu melt-spun ribbons, Journal of Alloys and Compounds 776 (2019) 319-325.

[32] X.M. He, L.J. Rong, D.S. Yan, Y.Y. Li, Temperature memory effect of Ni47Ti44Nb9 wide hysteresis shape memory alloy, Scripta Materialia 53(12) (2005) 1411-1415.

[33] C. Tang, T.X. Wang, W.M. Huang, L. Sun, X.Y. Gao, Temperature sensors based on the temperature memory effect in shape memory alloys to check minor over-heating, Sensors and Actuators A: Physical 238 (2016) 337-343.

[34] J. López-García, V. Sánchez-Alarcos, V. Recarte, J.A. Rodríguez-Velamazán, I. Unzueta, J.A. García, F. Plazaola, P. La Roca, J.I. Pérez-Landazábal, Effect of high-energy ball-milling on the magnetostructural properties of a Ni45Co5Mn35Sn15 alloy, Journal of Alloys and Compounds 858 (2021) 158350.

[35] H. Zheng, D. Wu, S. Xue, J. Frenzel, G. Eggeler, Q. Zhai, Martensitic transformation in rapidly solidified Heusler Ni49Mn39Sn12 ribbons, Acta Materialia 59(14) (2011) 5692-5699.

[36] Z. Wang, X. Zu, Y. Huo, Effect of heating/cooling rate on the transformation temperatures in TiNiCu shape memory alloys, Thermochimica Acta 436(1-2) (2005) 153-155.

[37] A.F. Manchón-Gordón, J.J. Ipus, J.S. Blázquez, C.F. Conde, A. Conde, P. Svec, Study of the kinetics and products of the devitrification process of mechanically amorphized Fe70Zr30 alloy, Journal of Alloys and Compounds 825 (2020) 154021.

[38] Y. Huo, X. Zu, On the Three Phase Mixtures in Martensitic Transformations of Shape Memory Alloys: Thermodynamical Modeling and Characteristic Temperatures, Continuum Mechanics and Thermodynamics 10(3) (1998) 179-188.